\documentclass[12pt]{article}
\usepackage{amssymb,amsmath,amsthm,latexsym,amsxtra,graphicx,appendix,epstopdf,feynmf,hyperref,setspace,fix-cm,color,cite,subfig}
\numberwithin{equation}{section}
\onehalfspace
\begin{document} 
\begin{titlepage}
\hbox to \hsize{\hspace*{0 cm}\hbox{\tt }\hss
    \hbox{\small{\tt }}}
\vspace{-1cm}
\hspace{12.4cm}BRX-TH-677

\vspace{2.5cm}
\centerline{\Large{\bf The $a$-theorem for the four-dimensional gauged vector model}}
\vspace{1 cm}
 \centerline{\large 
 {Howard J. Schnitzer\footnote{schnitzr@brandeis.edu}\, and Ida G. Zadeh\footnote{zadeh@brandeis.edu}}}
\vspace{0.3cm}
\centerline{\it Martin Fisher School of Physics, Brandeis University, Waltham, Massachusetts, 02454, USA}
\vspace{1 cm}
\begin{abstract}
The discussion of renormalization group flows in four-dimensional conformal field theories has recently focused on the $a$-anomaly. It has recently been shown that there is a monotonic decreasing function  which interpolates between the ultraviolet and infrared fixed points such that $\Delta a=a_{\rm UV}-a_{\rm IR}>0$. The analysis has been extended to weakly relevant and marginal deformations, though there are few explicit examples involving interacting theories. In this paper we examine the $a$-theorem in the context of the gauged vector model which couples the usual vector model to the Banks-Zaks model. We consider the model to leading order in the $1/N$ expansion, all orders in the coupling constant $\lambda$, and to second order in $g^2$. The model has both an IR and UV fixed point, and satisfies $\Delta a>0$.
\end{abstract}
\end{titlepage}
\tableofcontents
\setcounter{footnote}{0}
\newcommand{\be}{\begin{equation}}
\newcommand{\ee}{\end{equation}}
\newcommand{\bea}{\begin{eqnarray}}
\newcommand{\eea}{\end{eqnarray}}
\newcommand{\bb}{\mathbb}
\newcommand{\cl}{\mathcal}
\newcommand{\pr}{\partial}
\def\e{{\rm e}}
\def\cO{{\cal O}} 
\def\tx{\tilde{x}}
\def\half{{\textstyle{1 \over 2}}}
\def\eqn#1{eq.~(\ref{#1})} 
\def\Eqn#1{Equation~(\ref{#1})}
\def\eqns#1#2{eqs.~(\ref{#1}) and~(\ref{#2})}
\def\Eqns#1#2{Eqs.~(\ref{#1}) and~(\ref{#2})}
\section{Introduction}\label{I}
A persistent issue in quantum field theory (QFT) which has received considerable attention is whether renormalization group (RG) flows are reversible. That is, if there are conformal field theories (CFT) A and B such that there is a flow of 
A $\to$ B, can one also flow from B$ \to$ A? In two dimensions there exists a monotonically decreasing $c$-function, which connects ultraviolet (UV) and infrared (IR) CFT's which is known to be irreversible, such that the number of degrees of freedom decreases as one flows from the UV to the IR CFT's \cite{Zamolodchikov:1986gt}.

Similarly the same question for four-dimensional QFT's is of long-standing interest. A generalization of the two-dimensional discussion uses the fact that the trace of the stress-energy tensor $T_{\mu\nu}$ in a four-dimensional CFT in curved space has $T_\mu^{\;\mu}\ne0$ due to trace anomalies $a$ and $c$, where \cite{Deser:1993yx}
\be
T_\mu^{\;\mu}=aE_4-cW_{\mu\nu\rho\sigma}^2
\ee
with $E_4$ the Euler density and $W_{\mu\nu\rho\sigma}^2$ the square of the Weyl tensor. In this context, Cardy \cite{Cardy:1988cwa} conjectured that 
\be\label{a}
a_{\rm UV}>a_{\rm IR}
\ee
for flows from a UV CFT to an IR CFT. This was originally studied in a number of examples \cite{Osborn:1989td,Jack:1990eb,Schwimmer:2010za}. Holographic models of the $a$-theorem have also been studied in the context of the AdS/CFT correspondence and have provided proof of Cardy's conjecture for these models \cite{Myers:2010xs,Myers:2010tj}.

More recently Komargodski and Schwimmer \cite{Komargodski:2011vj} proved (\ref{a}) for all unitary RG flows such that there is a monotonic decreasing function which interpolates between $a_{\rm UV}$ and $a_{\rm IR}$. The theorem has been exemplified by several variants of non-interacting theories in various spacetime dimensions \cite{Luty:2012ww,Elvang:2012st,Elvang:2012yc}. There are however few known applications of the theorem to theories with interactions. The analysis of \cite{Komargodski:2011vj} was amplified further by Komargodski \cite{Komargodski:2011xv} to cases with weakly relevant deformations and to marginal deformations. An example of marginal deformation is the Banks-Zaks model \cite{Banks:1981nn}, for which case \cite{Komargodski:2011xv} computes $(\Delta a)_{\mathrm{BZ}}$ and shows that it satisfies (\ref{a}) in the flow from the UV to the IR fixed point.

Another interesting example of marginal deformations which we discuss in this paper is the gauged vector model in four dimensions \cite{Rhedin:1998gx,Olmsted:1996na} which couples the Banks-Zaks model to the usual vector model \cite{Schnitzer:1974ji,Schnitzer:1974ue,Dolan:1973qd,Root:1974zr,Coleman:1974jh,Kobayashi:1975ev,Schnitzer:1976aq,Abbott:1975bn}. The gauged vector model is reviewed in detail in section \ref{II}. The model has a scalar field $\phi^a$ and $N_f$ fermions $\psi^a$, both in fundamental representation of $U(N)$, with $\lambda(\phi^2)^2$ coupling, together with $U(N)$ Yang-Mills gauge fields coupled to the matter sectors. The calculations are carried out to leading order in $1/N$, to fourth order in the gauge coupling constant $g$, and all orders in $\lambda$. The couplings and $N_f/N$ can be chosen so that the theory is asymptotically free, and has a Banks-Zaks type marginal fixed point in the IR in both $(\lambda,g^2)$. With suitably chosen initial values of $(\lambda,g^2)$ and $N_f/N$, there is a flow from UV $\to$ IR fixed points. We will show, to the order which we compute, that using the methods of \cite{Komargodski:2011xv} the $a$-anomaly satisfies
\bea\label{deltaa}
\Delta a&=&a_{\rm UV}-a_{\rm IR}\\\nonumber
&=&(\Delta a)_{\mathrm{gauge}}+(\Delta a)_{\mathrm{scalar}}+(\Delta a)_{\mathrm{gauge-scalar}}>0.
\eea
If $(\lambda,g^2)$ do not lie in the range required for the existence of the IR fixed point, then either $(\lambda/g^2)\to\infty$ with $\Delta a\to+\infty$, or $g^2\to-\infty$, which is not a consistent gauge theory.

In section \ref{II} we review the gauged vector model, with RG flows of the coupling constants $(\lambda,g^2)$ and the conditions consistent (or not) for UV $\to$ IR fixed points. In section \ref{III} we use the Komargodski strategy for marginal deformations to compute $\Delta a$ for the gauged vector model, which will also expose the pathlogies described in section \ref{II}. A summary of the results appears in section \ref{V}.
\section{The gauged vector model}\label{II}
\subsection{The model}\label{II-i}
We consider the gauged $U(N)$ vector model in four spacetime dimensions \cite{Rhedin:1998gx,Olmsted:1996na}. The theory has gauged complex scalar fields, transforming in the fundamental representation of the $U(N)$ symmetry group, along with $N_f$ fermions which are also in the fundamental representation. The renormalized Lagrangian density of the theory is given by \cite{Rhedin:1998gx,Olmsted:1996na}
\bea\label{L}
\cl{L}\!\!\!&=&\!\!\!N\,\bigg\{|\pr_\mu\phi+iA_\mu\phi|^2+\frac{1}{2\lambda}\chi^2-\frac{\mu^2}{\lambda}\chi-\chi|\phi|^2+\nonumber\\
&&\quad+\sum_{j=1}{N_f}\left(\bar{\psi}_j\,\gamma\cdot D\psi_j\right)-\frac1{4\,g^2}\mathrm{Tr}\left(F_{\mu\nu}F^{\mu\nu}\right)\bigg\},
\eea
where $g$ and $\lambda$ are the gauge and scalar coupling constants, respectively, $\mu^2$ is the scalar mass, and $D$ is the covariant derivative. $A_\mu$ is the $U(N)$ gauge field transforming in the adjoint representation. $F_{\mu\nu}$ is the gauge field strength and $\chi$ is a Lagrange multiplier which is a singlet of $U(N)$. One can eliminate $\chi$ using the Euler-Lagrange equation in lowest order, which gives $\chi=|\phi|^2+2\mu^2$, and recover the $\lambda|\phi|^4$ interaction term and the mass term for the scalar field. In what follows we set $\mu^2/\lambda=0$.\footnote{Since $\beta_{\mu^2/\lambda}=[2\delta-\frac{6g^2}{16\pi^2}+\mathcal{O}(g^4)](\frac{\mu^2}{\lambda})$, it is consistent to do so where $d=4-\delta$ \cite{Olmsted:1996na}.}

The fields and the coupling constants in (\ref{L}) have been rescaled such that there is an overall factor of $N$ multiplying the Lagrangian density as follows: $\phi\to\sqrt{N}\phi$, $A_\mu\to\sqrt{N}A_\mu$, $g^2\to N^{-1}g^2$, and $\lambda\to N^{-1}\lambda$. Further, the gauge field is also rescaled according to $A_\mu\to g^{-1}A_\mu$. Then $1/N$ is an appropriate expansion parameter for the large-$N$ limit. Note that the Yukawa coupling between scalars and fermions is absent because both fields transform in the fundamental representation of $U(N)$.

We will present and evaluate the RG equations for the coupling constants for the purpose of computing the change in the $a$-anomaly for the RG flow from the UV to the IR fixed point. The details of the computation of $\Delta a$ are described in section \ref{III}. In the remaining part of this section we review the analysis of the renormalization group flow in the large $N$ limit of the $U(N)$ gauged vector model, and present the solutions to these equations.
\subsection{RG equations}\label{II-ii}
The RG equations satisfied by coupling constants in a renormalizable quantum field theory with scalar, fermion, and gauge fields and quartic scalar couplings have been derived in \cite{Machacek:1983tz,Machacek:1983fi,Machacek:1984zw} to two-loop order and to three-loop order in \cite{Pickering:2001aq} for fields transforming in different representations of a general gauge group. Using results of \cite{Machacek:1983tz,Machacek:1983fi,Machacek:1984zw}, $\beta$-functions for the scalar and gauge coupling constants in the $U(N)$ gauged vector model are evaluated in \cite{Rhedin:1998gx,Olmsted:1996na} in the large-$N$ limit, and are given by
\bea
\beta_g\!\!\!\!&=&\!\!\!\!\frac{dg}{d\log M}=-g\,\Big(b_0\,g^2+b_1\,g^4+\cdots\Big),\label{betag}\\ 
\beta_\lambda\!\!\!\!&=&\!\!\!\!\frac{d\lambda}{d\log M}=a_0\,\lambda^2-a_1\,g^2 \lambda+a_2\,g^4,\label{betalam}
\eea
where $M$ is an arbitrary mass scale, and the ellipsis correspond to higher order terms in $g^2$. The 3-loop gauge beta function is given in \cite{Pickering:2001aq}. It is noteworthy that the scalar coupling constant first appears in the 3-loop contribution to $\beta_g$, in a term of $\mathcal O(g^4\lambda)$ and one of $\mathcal O(g^2\lambda^2)$.

The $\beta$-functions (\ref{betag}) and (\ref{betalam}) are obtained at leading order in $1/N$, all orders in $\lambda$, and to second order in perturbative expansion in $g^2$. In the large $N$ limit and with the ratio $N_f/N$ fixed, the RG coefficients for the gauge coupling are given by \cite{Caswell:1974gg}
\bea\label{b0b1}
&&b_0={1 \over (4\pi)^2}\;\frac{4}{3}\left(\frac{11}2-\,{N_f\over  N} \right),\nonumber\\
&&b_1={1 \over (4\pi)^4}\;\frac{4}{3}\left(34-13\,{N_f\over  N} \right),
\eea
and the RG coefficients for the scalar coupling are \cite{Machacek:1983tz,Machacek:1983fi,Machacek:1984zw}
\bea\label{a0a1a2}
&&a_0={1 \over (4\pi)^2}\bigg[2 +16\,\left( g \over 4\pi\right)^2 + \cO(g^4)\bigg],\nonumber\\
&&a_1={1 \over (4\pi)^2}\bigg[12+{1\over 3} \left(256-40\,{N_f\over N} \right)\left( g \over 4\pi\right)^2+\cO(g^4)\bigg],\nonumber\\
&&a_2={1 \over (4\pi)^2}\bigg[6+{1\over 3} \left(304-64\,{N_f \over N} \right) \left( g \over 4\pi\right)^2+\cO(g^4)\bigg].
\eea
\subsection{Solutions to the RG equations}\label{II-iii}
In this subsection we present solutions to the RG equations. Following \cite{Rhedin:1998gx} we define two new variables $t$ and $x(t)$:
\bea
&&\;\;\;\;t\equiv \log M,\label{t}\\
&&x(t)\equiv g^2(t)\label{x(t)}.
\eea
The RG equation for the gauge coupling (\ref{betag}) then reads
\be\label{betax}
{dx\over dt}=-2b_0\,x^2(t)-2b_1\,x^3(t) +O(x^4(t)).
\ee
Following \cite{Rhedin:1998gx,Callaway:1988ya} we define two other variables
\bea
&&\;\;y\equiv{\lambda(t)\over g^2(t)},\label{y}\\
&&ds \equiv  x(t)\,dt.\label{sx}
\eea
The RG equation for the scalar coupling (\ref{betalam}) then reads
\be\label{betay}
{dy \over ds}=(a_0\,y^2-a_1\,y+a_2)+2\,(b_0+b_1\,x)\,y.
\ee
Note that the RG flow equation for the gauge coupling constant (\ref{betax}) does not depend on $y$ to second order in $x$ which we are considering. However, the RG flow of $y$ in (\ref{betay}) does depend on $x$. In order to solve the RG flow equation (\ref{betay}) we first evaluate $x(s)$. Inserting equation (\ref{sx}) in (\ref{betax}), the flow equation of $x$ reads
\be\label{betax-ii}
{dx\over ds}=-2b_0\,x(s)\Big(1-\frac{b_1}{b_0}\,x(s)\Big),
\ee
to second order in $x(s)$. Defining
\be\label{x0}
x(s=0)=g^2(s=0)\equiv x_0,
\ee
we integrate equation (\ref{betax-ii}) from $x(s=0)=x_0$ to $x(s>0)$, and find
\be\label{xofs-long}
-2b_0\int_{0}^{s}ds=\int_{x_0}^{x}\frac{dx}{1-\frac{b_1}{b_0}\,x},
\ee
and obtain
\be\label{xofs}
x(s)=\Bigg[\left(\frac{1}{x_0}-\frac{b_1}{b_0}\right)e^{b_0s}+\frac{b_1}{b_0}\Bigg]^{-1}.
\ee
Equations (\ref{sx}) and (\ref{xofs}) relate $t$ to $s$, as follows:
\be\label{tofs}
t=\frac{1}{b_0}\Bigg[\Big(\frac{1}{x_0}-\frac{b_1}{b_0}\Big)\;\Big(e^{b_0s}-1\Big)+b_1\,s\Bigg],
\ee
where $t$ is a single-valued function of $s$, with $s=0$ for $t=0$, and $s\to\pm\infty$ as $t\to\pm\infty$.

The RG flow equation (\ref{betay}) can be written as
\be\label{betay-ii}
\frac{dy(s)}{ds}=a_0\,\Big(y(s)-y_-(s)\Big)\,\Big(y(s)-y_+(s)\Big),
\ee
where
\be\label{ypm}
y_{\pm}(s)=\left(\frac{a_1(s)-b_0+b_1\,x(s)}{2a_0(s)}\right)\pm\sqrt{\left(\frac{a_1(s)-b_0+b_1\,x(s)}{2a_0(s)}\right)^2-\frac{a_2(s)}{a_0(s)}},
\ee
and $y_{\pm}(s)>0$ for all values of $s$. Equation (\ref{betay-ii}) is then solved using $x(s)$ derived in (\ref{xofs}) and the RG flow coefficients given in (\ref{b0b1}) and (\ref{a0a1a2}).

In order for the theory to be asymptotically free in the UV regime the following constraints are required to be satisfied \cite{Rhedin:1998gx,Olmsted:1996na}
\bea\label{asfree}
\frac{N_f}{N}\!\!\!&<&\!\!\!\frac{11}2\nonumber\\
\frac{\lambda}{g^2}\!\!\!&<&\!\!\!\frac23\left(\frac{N_f}{N}-1\right)+\sqrt{\frac49\left(\frac{N_f}{N}-1\right)^2-3}.
\eea
The theory has a Banks-Zaks fixed point for the gauge coupling constant in the IR regime when the ratio $N_f/N$ is the range \cite{Banks:1981nn}
\be\label{bz}
\frac{34}{13}<\frac{N_f}N<\frac{11}2.
\ee
The value of the gauge coupling at this fixed point is
\be\label{g*}
\left(\frac{g_*}{4\pi}\right)^2=\frac{1}{13}\;\frac{\frac{11}2-\frac{N_F}N}{\frac{N_F}N-\frac{34}{13}},
\ee
It is shown in \cite{Rhedin:1998gx} that for the ratio $N_f/N$ in the above range, there exist fixed points for massless scalar coupling constants in the IR limit. In addition, the reality condition of the coupling constants puts a more stringent constraint on the ratio then (\ref{bz}), i.e.,
\be\label{bz-ii}
\frac32\sqrt3+1\approx3.598<\frac{N_f}N<\frac{11}2.
\ee

\begin{figure}[!ht] 
  \subfloat[$N_f/N=4$\label{fig1a}]{%
    \includegraphics[width=0.5\textwidth]{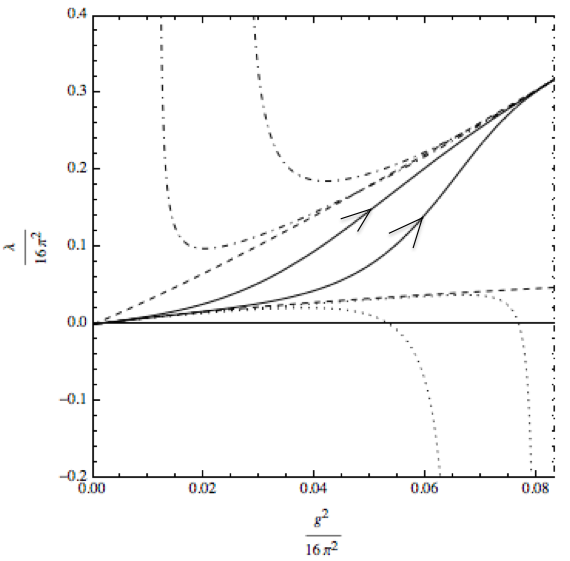} 
  } 
  \hfill 
  \subfloat[$N_f/N=5$\label{fig1b}]{%
    \includegraphics[width=0.5\textwidth]{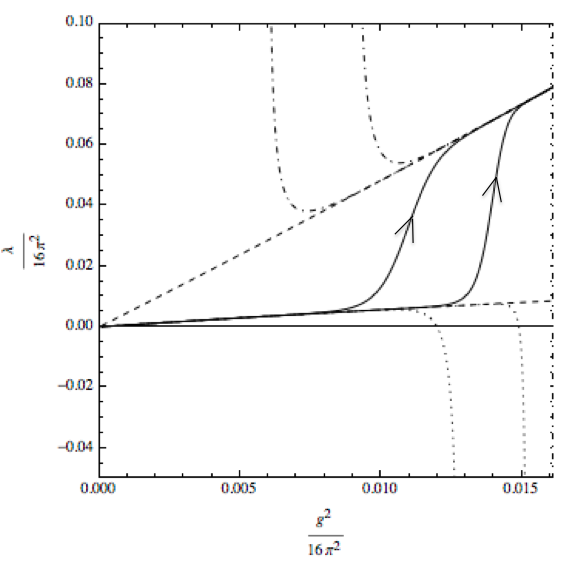}
  } 
  \caption{RG group flow for the gauged $U(N)$ vector model for two values of $N_f/N=4$ (\ref{fig1a}) and $N_f/N=5$ (\ref{fig1b}). Solid lines inside the wedge are RG flows starting from the UV fixed point at the lower left corner of the wedge and ending at the IR fixed point at the upper right corner. Different lines correspond to different values of the initial slope $y_{\rm ini}(s_{\rm ini})$. Note the sensitivity of the RG flows to the initial conditions in figure \ref{fig1b}.
} 
\label{fig1}
\end{figure}

The RG flow in the gauged $U(N)$ model is determined by solving equations $(\ref{betax-ii})$ and $(\ref{betay-ii})$. Appropriate values of the coupling constants, which are non-negative and which yield asymptotic freedom, are located inside a wedge in the space of the coupling constants restricted by the two curves $y_{\pm}(s)$ defined in (\ref{ypm})\cite{Rhedin:1998gx}. If the value of $y(s)$ is greater than $y_+(s)$, then the flow diverges to infinity in the UV limit: $y(s\to+\infty)\to+\infty$. On the other hand, if $y(s)<y_-(s)$, then the flow diverges to negative infinity in the IR regime $y(s\to-\infty)\to-\infty$. RG flows inside the wedge of physical values of $g^2$ and $\lambda$ start from $(g^2,\lambda)=(0,0)$ in the UV with the initial value of $y_{\rm ini}(s_{\rm ini})$ restricted by $y_-(s_{\rm ini})<y_{\rm ini}(s_{\rm ini})<y_+(s_{\rm ini})$, where $s_{\rm ini}\to+\infty$ at the UV fixed point. These flows end at the IR fixed point with $(g^2,\lambda)=(g_*^2,\lambda_*)$, where $\lambda_*=g_*^2\,y_+(-\infty)$. Figures \ref{fig1a} and \ref{fig1b} show RG flows for two different choices of $N_f/N=4$ and $N_f/N=5$, respectively. In each plot, $y_\pm(s)$ forms the wedge of allowed values of the coupling constants as represented by two dashed lines. The slopes of the dashed lines are $y_\pm(s)$ with $y_+(s)>y_-(s)$. Solid lines inside the wedge represent RG flows from UV to IR fixed points. Different curves correspond to different values of the initial condition $y_{\rm ini}(s_{\rm ini})$. Dotted-dashed lines above the wedge correspond to flows with $y(s)>y_+(s)$ and dotted lines below the wedge are flows with $y(s)<y_-(s)$. The right vertical side of each frame is located at $g_*^2$ derived in (\ref{g*}).
\subsubsection{Perturbative limit}\label{II-iii-i}
Consider RG flows from the UV fixed point to an IR fixed point such that the IR fixed point is reached perturbatively in the space of coupling constants. This allows us to use perturbation theory to analyze the RG flow and compute the change in the $a$-anomaly. Starting from the UV fixed point where coupling constants vanish, the IR fixed point is obtained for parametrically small values of the coupling constants. Consider equations (\ref{betag}) and (\ref{b0b1}) for the gauge coupling constant, with the number of fermions $N_f\to\infty$ such that $N_f/N$ approaches $11/2$ from below. Define 
\begin{equation}\label{eps}
\epsilon\equiv\frac{11}{2}-\frac{N_f}{N},\qquad\qquad\epsilon\ll1.
\end{equation}
The gauge RG coefficients (\ref{b0b1}) can be written in terms of $\epsilon$ as
\bea
&&b_0=\frac1{(4\pi)^2}\,\frac43\,\epsilon,\label{b0-ii}\\
&&b_1=\frac{-2}{(4\pi)^4}\left(25-\frac{26}{3}\,\epsilon\right).\label{b1-ii}
\eea
The scalar RG coefficients (\ref{a0a1a2}) in the limit (\ref{eps}) read
\bea\label{a0a1a2-eps}
&&a_0={1 \over (4\pi)^2}\bigg[2 +16\left(\frac{g}{4\pi}\right)^2\bigg],\nonumber\\
&&a_1={1 \over (4\pi)^2}\bigg[12+\left(\frac{g}{4\pi}\right)^2\left(12+\frac{40}3\epsilon\right)\bigg],\nonumber\\
&&a_2={1 \over (4\pi)^2}\bigg[6+\left(\frac{g}{4\pi}\right)^2\left(-48+64\epsilon\right)\bigg].
\eea

Using equation (\ref{betax-ii}), we find that the IR fixed point for the gauge coupling constant $g_*^2\equiv x_*=x(s\to-\infty)$ is located at
\be\label{xstar}
x_*=-\frac{b_0}{b_1},
\ee
to second order in $x$. Inserting coefficients (\ref{b0-ii}) and (\ref{b1-ii}) in (\ref{xstar}) we have
\be\label{xstar-ii}
x_*=(4\pi)^2\bigg(\frac{2}{75}\epsilon+\frac{52}{5625}\epsilon^2+O(\epsilon^3)\bigg).
\ee
Next we use (\ref{xstar-ii}) and write the scalar coefficients $a_0$, $a_1$, and $a_2$ (\ref{a0a1a2}) in terms of a perturbative expansion in $\epsilon$. We obtain
\bea\label{a0a1a2-iii}
&&a_0={1 \over (4\pi)^2}\bigg[2 +\frac{32}{75}\,\left(\frac{x(s)}{x_*}\right)\,\epsilon+\frac{832}{5625}\,\left(\frac{x(s)}{x_*}\right)\,\epsilon^2+O(\epsilon^3)\bigg],\nonumber\\
&&a_1={1 \over (4\pi)^2}\bigg[12+\frac{8}{25}\,\left(\frac{x(s)}{x_*}\right)\,\epsilon+\frac{2624}{5625}\,\left(\frac{x(s)}{x_*}\right)\,\epsilon^2+O(\epsilon^3)\bigg],\nonumber\\
&&a_2={1 \over (4\pi)^2}\bigg[6-\frac{32}{75}\,\left(\frac{x(s)}{x_*}\right)\,\epsilon+\frac{2368}{5625}\,\left(\frac{x(s)}{x_*}\right)\,\epsilon^2+O(\epsilon^3)\bigg].
\eea
The beta function for $y(s)$ in (\ref{betay}) is then written in the perturbative limit using the above relations. The wedge functions $y_\pm(s)$ (\ref{betay-ii}) in this limit are of the form
\bea\label{ypm-ii}
y_\pm(s)&=& (3\pm\sqrt6)+\left[-\frac13\left(2\pm\sqrt6\right)+\frac{4}{225}\left(6\pm7\sqrt6\right)\,\frac{x(s)}{x_*} \right]\,\epsilon+\nonumber\\
&+&\left[\mp\frac{\sqrt6}{54}+\frac{4}{5626}\left(52\pm69\sqrt6\right)\;\frac{x(s)}{x_*} -\frac{8}{50625}\left(144\pm211\sqrt6\right)\;\left(\frac{x(s)}{x_*}\right)^2 \right]\,\epsilon^2+\nonumber\\
&+&O(\epsilon^3).
\eea
At the IR fixed point $x(s\to-\infty)=x_*$ and
\be\label{yp-ir}
y_\pm(s\to-\infty)=(3\pm\sqrt6)-\frac1{225}\left(126\pm47\sqrt6\right)\epsilon+
\frac{\left(1440\mp483\sqrt6\right)}{101250}\,\epsilon^2+O(\epsilon^3).
\ee

\begin{figure}[!ht] 
  \subfloat[$\epsilon=0.1$\label{fig2a}]{%
    \includegraphics[width=0.51\textwidth]{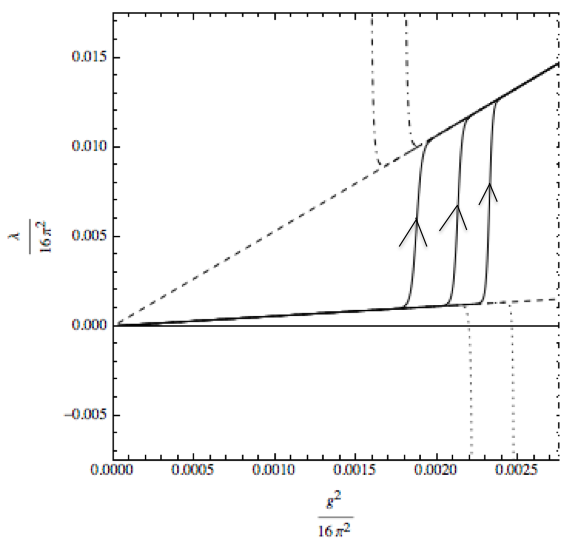} 
  } 
  \hfill 
  \subfloat[$\epsilon=0.01$\label{fig2b}]{%
    \includegraphics[width=0.51\textwidth]{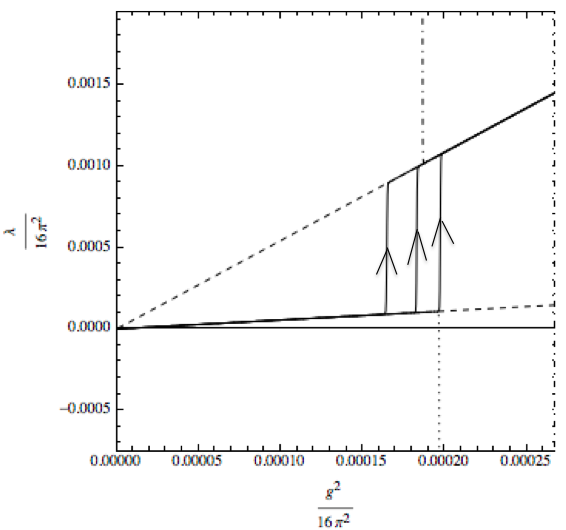} 
  } 
  \caption{RG group flow for the gauged $U(N)$ vector model in the perturbative limit (\ref{eps}) for two choices of $\epsilon=0.1$ (\ref{fig2a}) and $\epsilon=0.01$ (\ref{fig2a}). The line conventions are the same as those in figure \ref{fig1}. Note again the sensitivity of the RG flows to the initial conditions in both panels. Note also that in figure \ref{fig1} the values of $\epsilon$ are $1/2$ and $3/2$ which are not appropriate for the methods of ref. \cite{Komargodski:2011xv}.} 
\label{fig2}
\end{figure}

In the limit $\epsilon\ll1$, the leading $\epsilon$ behaviour of the gauge coupling constant (\ref{xstar-ii}) is
\be\label{x*-ii}
\frac{g_*^2}{(4\pi)^2}=\frac2{75}\epsilon.
\ee
The value of the scalar coupling constant is
\be\label{lambda*}
\frac{\lambda_*}{(4\pi)^2}=\frac{g_*^2}{(4\pi)^2}\;y_+(-\infty)\approx5.449\,\frac2{75}\,\epsilon,
\ee
to leading order in $\epsilon$. The smallness of $\epsilon$ (\ref{eps}) justifies parametrically small values of both the gauge and scalar coupling constants, which makes them suitable for performing perturbation theory. Figure \ref{fig2} plots the RG flow in the perturbative limit for the two choices of $\epsilon=0.1$ (\ref{fig2a}) and $\epsilon=0.01$ (\ref{fig2b}). Values of the gauge and scalar coupling constants are found at the IR fixed point using the vertical line on the right side of the frame. Note the sensitivity to the initial conditions in the two plots in Figure \ref{fig2}.
\section{The $a$-theorem for marginal perturbations}\label{III}
\subsection{Strategy}\label{III-i}
Komargodski \cite{Komargodski:2011xv} extended the discussion of \cite{Komargodski:2011vj} to the discussion of weakly relevant flows and to perturbation by marginal operators by promoting coupling constants to a function of spacetime. One considers the effective theory at mass-scale $M$ along the flow from a UV to an IR fixed point by writing the coupling constants in terms of $M$ and a compensator $F=\exp\tau$. In our case the couplings of the gauged vector model are modified by
\bea
g^2&\to&g^2(FM)=g^2(e^\tau M),\nonumber\\
\lambda&\to&\lambda(FM)=\lambda(e^\tau M),
\eea
which renders the theory conformal. One extracts the dependence on $\tau$ by expanding the Lagrangian (2.1) around $\tau=0$, and focusing on the leading non-trivial terms, which are quadratic in $\tau$. As shown in \cite{Komargodski:2011xv}, the expansion in $\tau$ corresponds to an expansion in the ``distance" from the fixed point in coupling constant space, so that corrections to the leading non-trivial contribution are suppressed. This strategy is appropriate to the gauged vector model described in section \ref{II}.
\subsection{Application to the gauged vector model}\label{III-ii}
In order to compute $\Delta a$ we evaluate the Euclidean path integral and determine its $\tau$-dependence. The coefficient of four-derivative terms in $\tau$ in the effective dilaton action gives the change in the $a$-anomaly \cite{Komargodski:2011xv}. In the gauged vector model we consider, the gauge and scalar coupling constants are parametrically small and the IR fixed point is reached perturbatively from the UV fixed point. The Euclidean path integral is then expanded perturbatively around the free theory. The strategy explained in section $\ref{III-i}$ can be implemented for (2.1), where the $\tau$-dependence is obtained by expanding the Lagrangian about $\tau=0$. This gives
\bea\label{lagiii}
N^{-1}\mathcal{L}&=&|\partial_\mu\phi+iA_\mu\phi|^2+\nonumber\\
&+&\frac12\,\frac1{\lambda(M)}\,\chi^2+\frac12\,\tau\,\beta_{\frac1{\lambda}}\,\chi^2+\frac14\,\tau^2\,\dot\beta_{\frac1{\lambda}}\,\chi^2+\cdots+\nonumber\\
&-&\chi|\phi|^2+i\sum_{i=1}^{N_f}(\bar\psi\,\gamma\!\cdot\!D\,\psi)+\nonumber\\
&-&\frac1{4g^2(M)}\mathrm{Tr}(F_{\mu\nu}F^{\mu\nu})-\frac1{4}\,\tau\,\beta_{\frac1{g^2}}\,\mathrm{Tr}(F_{\mu\nu}F^{\mu\nu})-\frac1{8}\,\tau^2\,\dot\beta_{\frac1{g^2}}\,\mathrm{Tr}(F_{\mu\nu}F^{\mu\nu})+\cdots,
\eea
where the ellipsis correspond to higher order terms in $\tau$ expansion{\footnote{Our gauge coupling constant $g$ differs by a factor of 2 from the coupling constant $g_{K}$ in \cite{Komargodski:2011xv} due to different conventions for the Lagrangian in our (2.1) as compared to equation (3.5) of \cite{Komargodski:2011xv}. The two coupling constants are related as $2g^2=g_K^2$}}. The $\beta$-functions in (\ref{lagiii}) are
\be\label{betas}
\beta_{\frac1\lambda}=-\frac1{\lambda^2}\beta_\lambda,\qquad\qquad\qquad\beta_{\frac1{g^2}}=-\frac2{g^3}\beta_g,
\ee
where $\beta_\lambda$ and $\beta_g$ are given explicitly in (2.2)$-$(2.5).

In the expansion of the functional integral, terms which are quadratic in $\tau$ yield four-derivative terms in the effective action to the leading order in $\epsilon$ \cite{Komargodski:2011xv}. These terms are of the form
\bea\label{tautau}
&&\frac1{8\lambda^4}(\beta_\lambda)^2\int d^4x\,d^4y\;\tau(x)\tau(y)\;\Big\langle\chi^2(x)\;\;\chi^2(y)\Big\rangle+\nonumber\\
&+&\frac1{8g^6}(\beta_g)^2\int d^4x\,d^4y\;\tau(x)\tau(y)\;\Big\langle\mathrm{Tr}(F_{\mu\nu}^2)(x)\;\;\mathrm{Tr}(F_{\mu\nu}^2)(y)\Big\rangle+\nonumber\\
&-&\frac14\,\frac1{\lambda^2}\,\frac1{g^3}\,\beta_\lambda\,\beta_g\int d^4x\,d^4y\;\tau(x)\tau(y)\;\Big\langle\chi^2(x)\;\;\mathrm{Tr}(F_{\mu\nu}^2)(y)\Big\rangle.
\eea
To obtain the result to leading non-trivial order in the coupling constants, we compute the correlators in (\ref{tautau}) to leading order, i.e., in free-field theory. The result is
\bea\label{corrs}
&&\Big\langle\chi^2(x)\;\;\chi^2(y)\Big\rangle=\frac{1}{256\,\pi^8}\,\frac{N^2\lambda^4}{(x-y)^8},\label{cc}\\
&&\Big\langle\mathrm{Tr}(F_{\mu\nu}^2)(x)\;\;\mathrm{Tr}(F_{\mu\nu}^2)(y)\Big\rangle=\frac{48}{\pi^4}\,\frac{N^2g^4}{(x-y)^8},\label{ff}\\
&&\Big\langle\chi^2(x)\;\;\mathrm{Tr}(F_{\mu\nu}^2)(y)\Big\rangle=0\label{cf}.
\eea
The two-point function (\ref{cf}) vanishes at leading order in $\epsilon$ in perturbation theory. That is, 
\be\label{mixedcorr}
\Big\langle\chi^2(x)\;\;\mathrm{Tr}F_{\mu\nu}^2(y)\Big\rangle\sim\Big[\big\langle\chi(x)\;\;F_{\mu\nu}^a(y)\big\rangle\Big]^2=0,
\ee
since $\chi(x)$ is a $U(N)$ singlet. The contribution to the effective action of the dilaton comes from the energy slice $d\log M$, which following \cite{Komargodski:2011xv} is of the form
\bea\label{dlogM}
\int d^4x\,d^4y\;\tau(x)\tau(y)\,\frac{1}{(x-y)^8}=\frac1{192}\Omega_3\int d^4x\;\tau(x)\,\Box^2\tau(x)\,d(\log M),
\eea
where $\Omega_{d-1}=2\pi^{d/2}/\Gamma(d/2)$. Inserting (\ref{cc})-(\ref{cf}) and (\ref{dlogM}) in (\ref{tautau}), the leading dilaton contribution reads
\bea\label{tboxt}
&&\;\;\;\frac{\Omega_3\,N^2}{192}\int d^4x\;\tau(x)\,\Box^2\tau(x)\,d(\log M)\Big\{\frac{1}{2048\,\pi^8}\,(\beta_\lambda)^2+\frac{6}{\pi^4\,g^2}\,(\beta_g)^2\Big\}\nonumber\\
&&=\frac{\Omega_3\,N^2}{192}\int d^4x\;\tau(x)\,\Box^2\tau(x)\,d(\log M)\times\nonumber\\
&&\times\Bigg\{\frac{1}{2048\,\pi^8}\,\bigg[a_0\,\lambda^2(M)-a_1\,g^2(M)\,\lambda(M)+a_2\,g^4(M)\bigg]^2+\nonumber\\
&&+\frac{6}{\pi^4\,g^2(M)}\bigg[b_0\,g^3(M)+b_1\,g^5(M)\bigg]^2\Bigg\}.
\eea
Observe from (\ref{betag}) and (\ref{b0b1}) that $\beta_g$ in the last line in (\ref{tboxt}) is independent of the scalar coupling constat $\lambda$. We first consider the contribution from $\beta_g$ and integrate over all energies to compute the change in the corresponding $a$-anomaly $(\Delta a)_{{\rm gauge}}$. Following \cite{Komargodski:2011xv}, we get
\be
(\Delta a)_{\mathrm{gauge}}=\frac12\,\frac{\Omega_3\,N^2}{192}\,\frac{6}{\pi^4}\int d(\log M)\;\frac{\Big(\beta_g(M)\Big)^2}{g^2(M)}=\frac{\Omega_3\,N^2}{64\,\pi^4}\int_{g_*}^0\frac{dg}{g^2}\,\beta_g.
\ee
Using equation (\ref{betag}) we obtain{\footnote{There is a factor of four difference with the $a$-anomaly computed in \cite{Komargodski:2011xv} due to the different conventions of $g^2$ described in footnote 3.}}
\be
(\Delta a)_{\mathrm{gauge}}=\frac{N^2}{900\,\pi^2}\,\epsilon^2>0\label{da-g-1}.
\ee
Similarly, we evaluate $(\Delta a)_{\mathrm{scalar}}$ from the terms proportional to $(\beta_\lambda)^2$ in (\ref{tboxt}). The result is
\be\label{da-s}
(\Delta a)_{\mathrm{scalar}}=\frac12\,\frac{\Omega_3\,N^2}{192}\frac{1}{2048\,\pi^8}\int d(\log M)\;\Big[a_0\,\lambda^2(M)-a_1\,g^2(M)\,\lambda(M)+a_2\,g^4(M)\Big]^2>0.\ee
Using (\ref{betalam}) the above equation reads
\be
(\Delta a)_{\mathrm{scalar}}=\frac12\,\frac{\Omega_3\,N^2}{192}\frac{1}{2048\,\pi^8}\int_{\lambda_*}^{0} d\lambda\;\beta_{\lambda}=\mathcal O(\epsilon^3).
\ee
This integral is evaluated numerically using discussions of section \ref{II-iii} and the result is of subleading order of $\mathcal O(\epsilon^3)$ compared to the leading gauge contribution (\ref{da-g-1}). The value of this subleading term $(\Delta a)_{\rm scalar}$ depends on the path of the RG flow inside the wedge, i.e. on the initial condition $y_{\rm ini}(s_{\rm ini})$, in the space of the coupling constants. At $\mathcal O(\epsilon^3)$, the contribution to $(\Delta a)_{\rm gauge}$ also depends on the scalar coupling constant because $\beta_g$ contains terms of $\mathcal O(g^4\lambda)$ and $\mathcal O(g^2\lambda^2)$ at the 3-loop level \cite{Pickering:2001aq}. The gauge contribution to the $a$-anomaly then depends on the path of the flow at the subleading order as well and this would presumably cancel the $\mathcal O(\epsilon^3)$ path dependence in (\ref{da-s}). There exist further contributions from higher order corrections to (\ref{tautau}) which yield additional subleading corrections in $\epsilon$. The detailed analysis of $\mathcal O(\epsilon^3)$ terms is beyond the scope of the present work. Note that for $(\Delta a)_{\rm scalar}$ to be of subleading $\epsilon^3$ order, the perturbative limit requirement $\epsilon\ll1$ has to be satisfied. However, note the very small coefficient in (\ref{da-s}) so that $\epsilon=0.1$ is adequate for our purposes.

The total change in the $a$-anomaly to the leading order in $\epsilon$ is thus 
\be\label{delta-a-tot}
(\Delta a)_{\rm tot}=(\Delta a)_{\rm gauge}+(\Delta a)_{\rm scalar}+(\Delta a)_{\rm gauge-scalar}=N^2\Big(\frac{1}{900\,\pi^2}\,\epsilon^2+O(\epsilon^3)\Big)>0,
\ee
where $(\Delta a)_{\rm gauge}=\mathcal O(\epsilon^2)$, $(\Delta a)_{\rm scalar}=\mathcal O(\epsilon^3)$, and $(\Delta a)_{\rm gauge-scalar}$ vanishes at the leading order in perturbation theory.

Observe in figure \ref{fig2}, and section \ref{II-iii}, that flows for $y(s)>y_+(s)$ grow in the UV to $y(s)|_{s\to+\infty}\to+\infty$, which implies that
\be
(\Delta a)_{\mathrm{scalar}}\to+\infty.
\ee
 For flows for which $y(s)<y_-(s)$, one evolves in the IR to $y(s)\to-\infty$, which is excluded as it leads to unphysical result $(\lambda/g^2)<0$. These pathologies restrict the application of (\ref{tboxt}) to allowed flows in figure \ref{fig2}.
\section{Conclusions}\label{V}
In this paper we extended the analysis of \cite{Komargodski:2011xv} to a more general Banks-Zaks type fixed point, i.e., the gauged vector model in four dimensions. For allowed flows, we find from the argument of \cite{Komargodski:2011xv}
\bea\label{deltaa-conc}
\Delta a&=&a_{\rm UV}-a_{\rm IR}\\\nonumber
&=&(\Delta a)_{\mathrm{gauge}}+(\Delta a)_{\mathrm{scalar}}+(\Delta a)_{\mathrm{gauge-scalar}}>0,
\eea
where $(\Delta a)_{\rm gauge}=\mathcal O(\epsilon^2)$, $(\Delta a)_{\rm scalar}=\mathcal O(\epsilon^3)$, and $(\Delta a)_{\rm gauge-scalar}$ vanishes to the leading order in perturbation theory. The expansion (\ref{lagiii}) corresponds to an expansion in the ``distance" from the UV to IR fixed points to the leading non-trivial order in the couplings. There are $\mathcal O(\epsilon^3)$ subleading contributions to the $a$-anomaly from both gauge and scalar coupling constants. The $\mathcal O(\epsilon^3)$ gauge contribution comes in at the three-loop level where $\beta_g$ contains terms of the form $g^6$, $g^4\lambda$, and $g^2\lambda^2$. Furthermore, there are additional subleading corrections to $\Delta a$ from higher order expansion in dilaton in (\ref{tautau}). The property (\ref{deltaa-conc}) presumably generalizes to other Banks-Zaks models. For example see \cite{Antipin:2013pya}.
Flows of $y(s)=\lambda(s)/g^2(s)$, with $y>y_+(s)$ grow in the UV to $y(s)\to\infty$, and thus $\Delta a=+\infty$. Similarly, flows for which $y(s)<y_-(s)$ as $y(s)\to-\infty$ are pathological in the IR.
\section*{Acknowledgements}
We wish to thank Steve Naculich for his participation in the early phases of this project. We are appreciative of Matt Headrick and Albion Lawrence  for helpful conversations and comments. HJS is supported in part by the DOE by grant DE-SC0009987. IGZ is supported in part by the DOE by grant DE-SC0009987 and in part by the NSF by grant PHY-1053842.
\bibliographystyle{utphys}
\bibliography{list}
\end{document}